 \documentstyle[epsfig]{article}

\twocolumn
\begin{document} 
\newcommand{\IGNFIG}[1]{{}}
\newcommand{\IGN}[1]{{}}
\title{\mbox{Self-organized pore formation and open-loop-control in semiconductor etching}}
%\titlerunning{Open-Loop-Control of Semiconductor Pores}
\author{Jens Christian Claussen$^{1,2}$, J\"urgen Carstensen$^{1}$, Marc Christophersen$^{1}$, Sergiu Langa$^{1}$,
and Helmut F\"oll$^{1}$ 
\\$^1$Chair for general materials science, CAU University of Kiel,
Kaiserstr. 2, 24143 Kiel, Germany\\
\texttt{http://www.tf.uni-kiel.de/matwis/amat/}
\\$^2$Theoretische Physik und Astrophysik, Universit\"at Kiel,
Leibnizstr. 15, 24098 Kiel, Germany\\
\texttt{http://www.theo-physik.uni-kiel.de/\~{ }claussen/} }
\date{{\sl Published in CHAOS (AIP) 13 (1), 217-224 (2003)}}
\maketitle
\begin{abstract}
Electrochemical etching of semiconductors, apart from many technical
applications, provides an interesting experimental setup for
self-or\-ga\-ni\-zed structure formation capable e.g. of regular, 
dia\-me\-ter-mo\-du\-la\-ted, and branching pores.
The underlying dynamical processes governing current transfer
and structure formation are described by the Current-Burst-Model: 
all dissolution processes are assumed to occur inhomogeneously in time 
and space as a Current Burst (CB); the properties and interactions between CB's 
are described by a number of material- and chemistry- dependent ingredients,
like passivation and aging of surfaces in different crystallographic orientations,
giving a qualitative understanding of resulting pore morphologies.
These morphologies cannot be influenced only by the current, 
by chemical, material and other etching conditions, but
also by an open-loop control, triggering the time scale
given by the oxide dissolution time. 
With this method, under conditions where only branching
pores occur, the additional signal hinders side pore formation
resulting in regular pores with modulated diameter.
\end{abstract}
%wc abstract.txt 
%      17     119     915 abstract.txt
%
%
%
\typeout{ Leading paragraph, to the non-specialized reader}
%
%
% \clearpage
\noindent \bf
\\
Silicon monocrystals are the most perfect material 
mankind has ever created
since they are essential for high integrated computer circuits.
Due to this high degree of perfection, 
i.~e. corrosion is not defect-driven, 
and ``ideal corrosion'' is possible:
Silicon and other semiconductors can be made porous by
electrochemical etching giving an outstanding variety of pore
sizes, from nanopores to macropores, and geometries,
including disordered nanoporous, dendritic-like sidebranching pores,
and pores with modulated diameter.
Abstracting from the details of the underlying electrochemical processes, 
the Current Burst Model together with the Aging Concept,
and accounting the interactions between Current Bursts, 
the generation of different pore geometries,
oscillations and synchronization phenomena can 
be explained including the percolation transition
to global oscillations.
Based on the time scale derived from the Aging Concept,
an open-loop control can be applied to suppress
sidebranching of pores
in a technologically relevant regime.
\rm

\section{Introduction}
\begin{figure}[thbp] \noindent
\epsfig{file=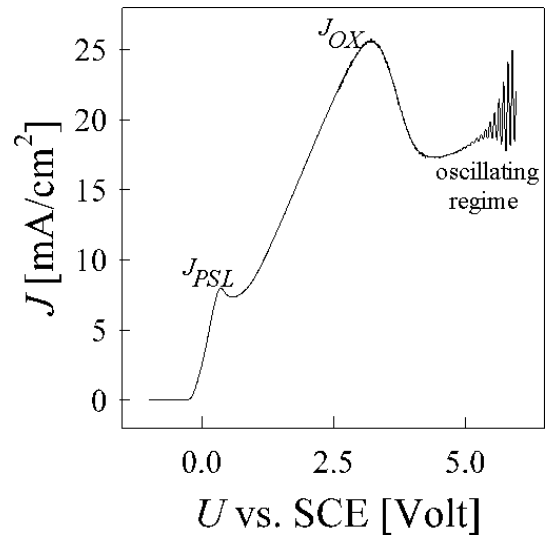,width=0.8\columnwidth}
\caption{ \label{fig_kennlinie}
The IV- characteristics of the silicon-hydrofluoric acid contact shows 
different phenomena from generation of a porous silicon layer (PSL), oxidation and
electropolishing (OX) and electrochemical oscillations at higher anodic bias.
}
\end{figure}

The solid - liquid junction of Silicon and HF - 
containing liquids exhibits a number of peculiar features, 
e.g. a very low density of surface states, 
i.e. an extremely well ``passivated'' interface 
\cite{yablonovich86}. 
If the junction is biased, the IV - characteristics 
(Fig.~\ref{fig_kennlinie}) in diluted 
HF is quite complicated and exhibits two current peaks and 
strong current- or voltage oscillations at large current 
densities (for reviews see \cite{foell91appl,smith92}).
These oscillations have been decribed quantitatively by the 
Current-Burst-Model
\cite{carstensen00matsciengb,carstensen98applphys,carstensen98sandiego,carstensen99electrochem}.

% Fig1 (kennlinie.eps)

%\IGNFIG{
%}%IGNFIG

Perhaps the most outstanding features are the many different 
kinds of pores - nanopores, mesopores, macropores, and so on 
- that form under a wide range of conditions in many HF containing 
electrolytes, including organic substances \cite{propst94,ponomarev98}. 
Despite of an intensive research triggered by the finding that 
nanoporous Si shows strong luminescence \cite{canham90,lehmann91}, 
neither the intricacies of the IV - characteristics nor the 
processes responsible for the formation of pores, including 
their rather peculiar dependence on the crystal orientation, 
are well understood.

Replacing Si by III/V-compounds, 
a variety of different pore morphologies can be 
etched; due to different properties of the
A- and B- surfaces one finds e.g. tetrahedron-shaped
pores instead of octahedron-shaped pores.
For an overview over recent results see  \cite{langa01theway}.
But again most of the phenomenon can be well understood within the 
framework of the Current-Burst-Model which seems to reflect a number of 
quite general properties of semiconductor electrochemistry.

%fig2 image49.eps

%
\section{Experimental Setup and Basic IV Characteristics}
%
%\IGNFIG{
\begin{figure}[thbp] \noindent
\epsfig{file=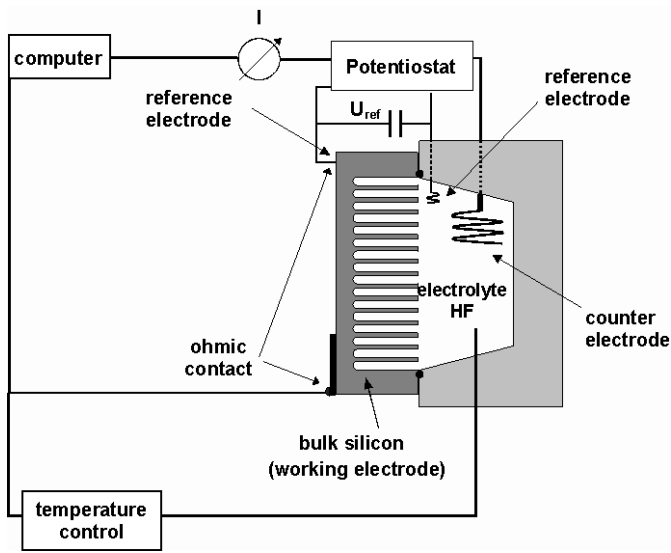,width=\columnwidth}
\caption{
 \label{fig_setup}
The experimental set-up used for electrochemical anodization of semiconductors.
}
\end{figure}
%}%IGNFIG

The basic setup is shown in Fig.~\ref{fig_setup}. 
Using a four electrode arrangement 
a po\-ten\-tio\-stat/gal\-va\-no\-stat 
is contacting the sample and the electrolyte, allowing for a 
well defind potential resp. current for the electrochemical 
dissolution reactions.
Since the  po\-ten\-tio\-stat/gal\-va\-no\-stat as well as the 
temperature are PC controlled,
all relevant etching parameters can be controlled in detail.
While the principal setup remains the same
in all experiments, backside contact,
front- and/or backside illumination and electrolyte pumping
can be varied as well as cell size (from under 0.3~cm up to wafers of 6~in)
and semiconductor material (Si, InP, GaAs, GaP)
including various doping levels and crystallographic orientations.
In addition, the electrolytes (e.g. HF, HCl, H$_2$SO$_4$) and their 
concentrations and temperature can be varied.
%

%\clearpage

\section{The Current-Burst Model}
\begin{figure}[ht]
 \noindent
\epsfig{file=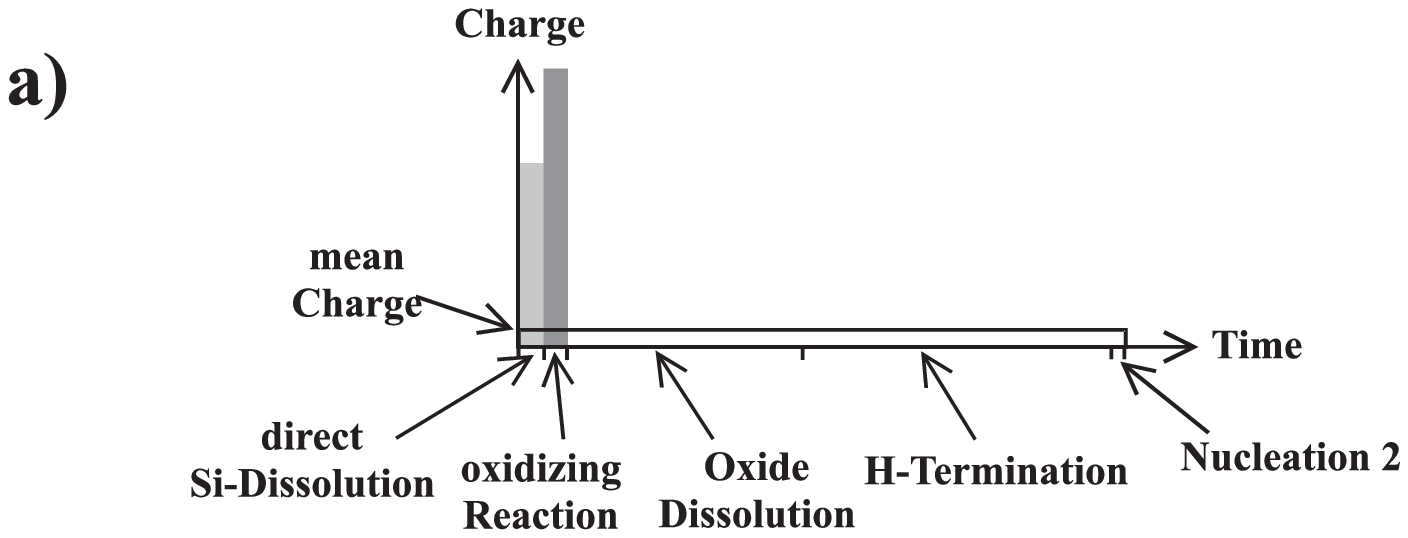,width=\columnwidth}
\\ \\
\epsfig{file=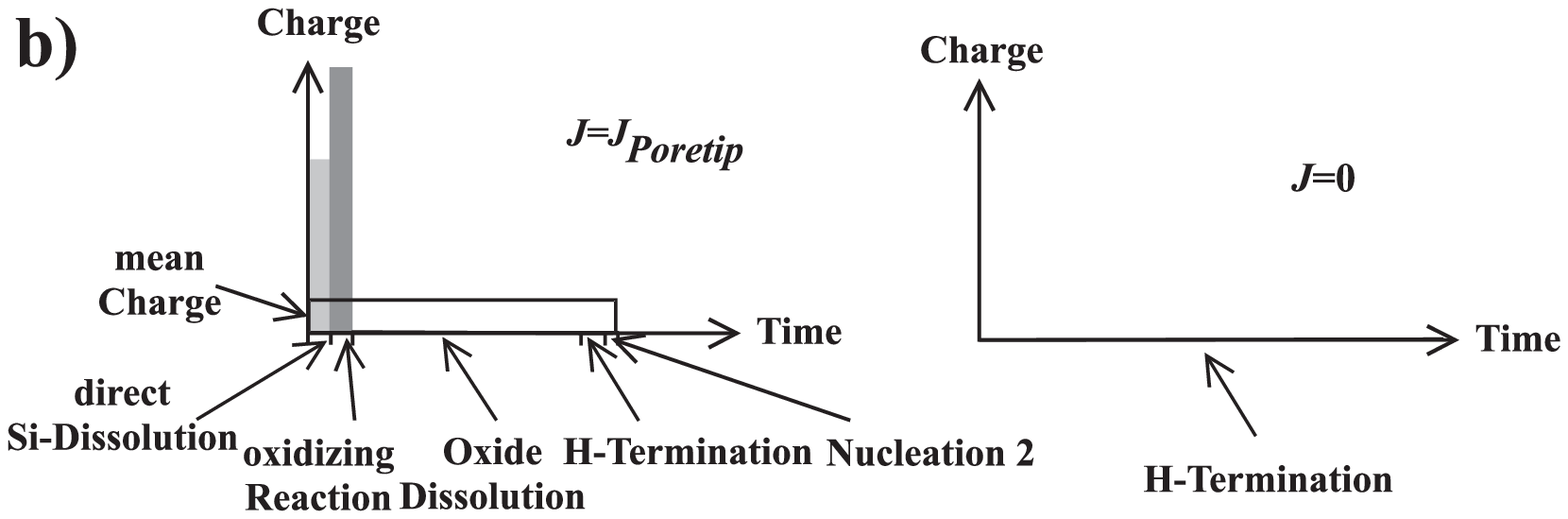,width=\columnwidth}
\\ \\
\epsfig{file=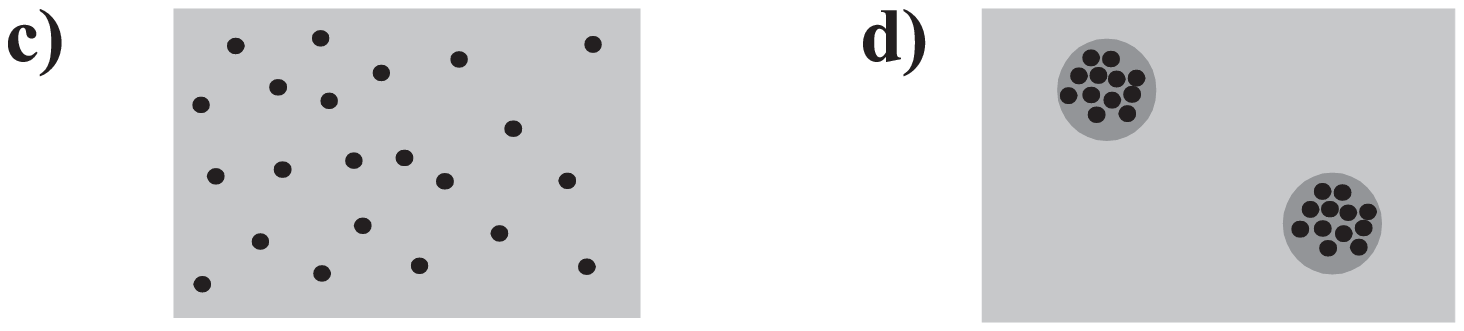,width=\columnwidth}
\caption{%
\label{fig_schem_charge_time_consuming}%
%\label{fig_applying_glob_curr}%
Applying a global current density smaller than the average current 
density in a current burst will either require very long time constants to 
keep the mean charge density passed in a burst small and to cover the 
surface completely with current lines (a), a statistical arrangement 
of current lines with the optimized (smaller) time constant of the system, 
leaving parts of the surface without current and creating nanopores (b, c), 
or induces a phase separation by rearranging the current lines in areas 
corresponding to macropores (d).
}
\end{figure}

The Current Burst Model \cite{carstensen00matsciengb,carstensen98applphys,carstensen98sandiego,foellarizona}
states that the dissolution 
mainly takes place on small spots in short events, 
starting with a direct Si-dissolution,
and possibly followed by an oxidizing reaction
(see Fig.~\ref{fig_schem_charge_time_consuming}).
%
%\IGNFIG{
%}%IGNFIG
%

After these two short processes, 
the oxide hump undergoes dissolution,
a time-consuming process which
ensures at the location of the current burst 
a dead-time of fixed length during which no new burst can start.
However, immediately after dissolution the 
Si surface has the highest reactivity,
resulting in a maximal probability of annother
current burst.
Due to H-termination the surface becomes passivated,
and the probability for bursts decays 
until it reaches the properties of a completely
passivated Si surface, comparable to the
situation before the nucleation of the first Current Burst.

This approach allows to connect the average current density $\overline{j}$ 
with a series of charge and time consuming processes
\begin{equation}
\overline{j} = \frac{\sum Q_i}{\sum {\tau}_i}. 
\label{cb_ratio}
\end{equation}
For most of these processes the dependence on the choice of the etching 
conditions is known; so by ``designing'' the electrochemical 
etching conditions (electrolyte, current density, applied voltage, 
temperature, illumination,...) the properties and interactions 
of the basic etching process can be tailored to create various pore
 morphologies.
In addition (\ref{cb_ratio}) directly couples the local average 
current density $\overline{j}$ to an intrinsic time constant 
$\sum {\tau}_{i}$. The influence of both, current density and intrinsic 
time constant, can be measured independently in experiments.
Independent measurements can be described with the
same set of parameters.
Within the current burst model pore formation from the nm to the $\mu$m 
range as illustrated by  
Fig.~\ref{fig_schem_charge_time_consuming}~c+d is very easy to 
describe because the separation of the dissolution processes and the 
surface passivation processes is already an immanent 
property of each dissolution cycle, i.e.{} current burst.  

%\clearpage

%\IGNFIG{
\begin{figure}[thbp] \noindent 
\epsfig{file=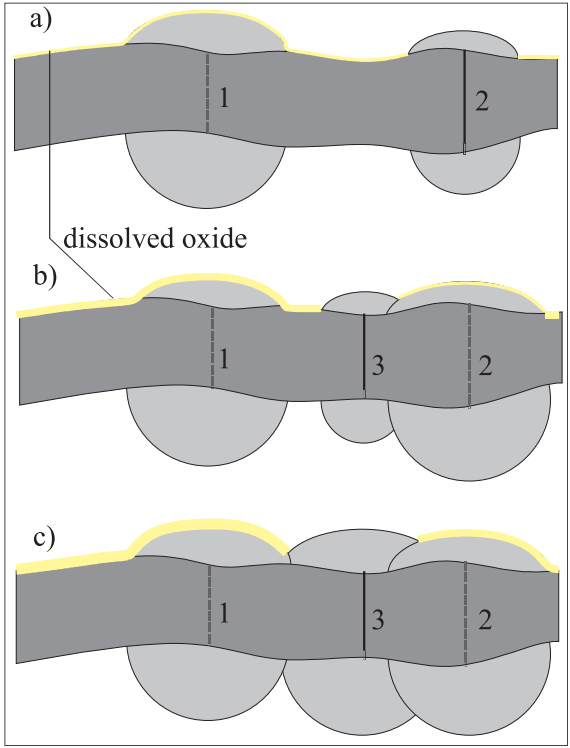,width=0.85\columnwidth}
\caption{ \label{fig_ionic_brkthru}
Schematic view of a ``ionic breakthrough''. 
Due to the oxidizing current a roughly semi-spherical 
oxide inclusion is growing from the tip of channel 1. 
This is the situation of extremely high current
(electrochemical oscillations).
}
\end{figure}
%}%IGNFIG

%\clearpage

\section{Smoothing effect of Oxide Layers and global current oscillations}
Depending on the regime in the IV-curve, 
Current Bursts and their interaction play
different roles: 
For extremely high currents, in the oscillation
regime, one has a permanent oxide coverage, and 
due to the high forcing of the system
current bursts are started at all 
locations where the oxide layer is 
sufficiently thin for a breakthrough
(Fig.~\ref{fig_ionic_brkthru}).
In this regime, a detailed Monte Carlo study
\cite{carstensen98applphys,carstensen98sandiego,carstensen99electrochem}
including the lateral interaction between
Current Bursts (overlap of the oxide humps of neighboring Current Bursts) 
has shown that the Current Burst Model
quantitatively can explain the experimental 
observations of globally oscillating etching current.
Due to a phase synchronization of neighboring Current Bursts oxide domains are formed. 
The size of this domains increases with increasing oxide generation/reduced oxide dissolution.
At a percolation point only one oxide domain exists on the sample surface with a 
synchronized cycle of oxide growth and dissolution, resulting in a
macroscopic
oscillation of the external current.
The size of the domains as well as the oscillation time can be controlled 
by the chemical parameters. Even without global oscillations for all regions of the IV-curve 
where oxide is formed one finds domains of synchronized oxide growth which define the length 
scale for the roughness of the electrochemically polished sample surface and thus lead 
to a smoothing of the surface.

%\clearpage

\section{Passivation Effects: The Aging Concept}
\begin{figure}[htbp] \noindent
\epsfig{file=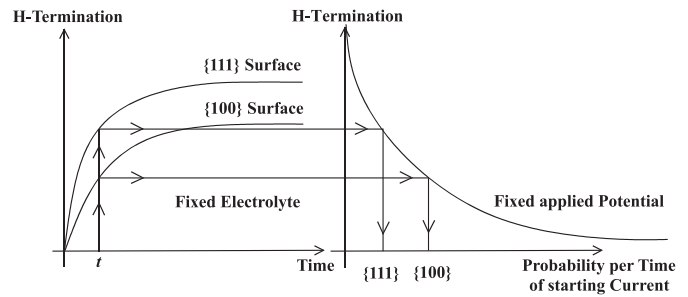,width=\columnwidth}
\caption{ \label{fig_h_termination}
Time constant and perfection of H-termination differ 
strongly for (100) and (111) Si surface orientation. 
This leads to an increased probability of local 
current nucleation on (100) surfaces.
}
\end{figure}

\begin{figure}[htbp]\noindent 
\epsfig{file=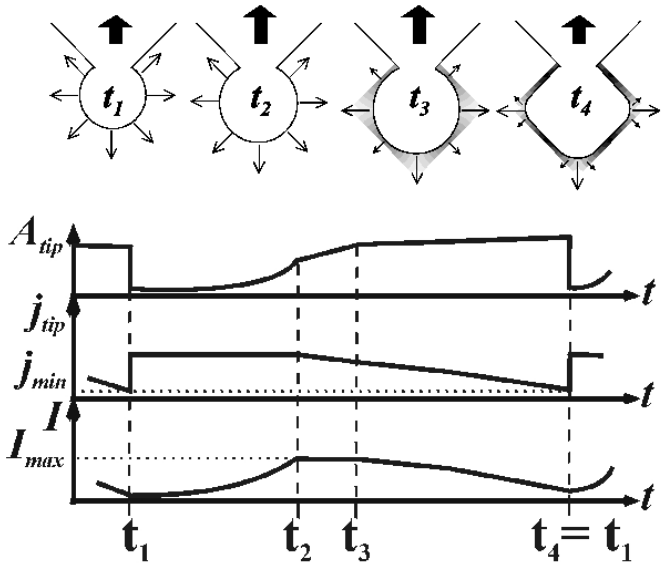,width=\columnwidth}
\caption{ 
\label{fig_cavities}
A consequence of the crystallographic dependence of the surface passivation in Fig.
\ref{fig_h_termination} 
is the growth of octahedra like pores. This pore growth is associated with
an oscillation of the current 
through this pore, even for a fixed externally applied potential.
}
\end{figure}

While at high current densities the semiconductor surface is completely covered with oxide at 
low current densities, most of the semiconductor surface will be in direct contact to the electrolyte.
It is well known \cite{yablonovich86} that after chemical dissolution the free surface is passivated, 
i.e. the density of surface states reduces as a function of time which increases the stability of the 
surface against further electrochemical attack. Schematically the perfection of the surface passivation 
and the resulting reduction of the probability for a chemical attack as a function of time are plotted 
in Fig. \ref{fig_h_termination}. For the example of silicon the speed and the perfection of 
passivation of the (111) crystallographic surface is larger than for the (100) surface. This selective 
aging of surfaces leads to a self amplifying dissolution of (100) surfaces (which will become pore tips) 
and a preferential passivation of (111) surfaces (which will become pore walls). Under optimized chemical 
conditions with an extremely large passivation difference between (111) and (100) surfaces a self 
organized growth of octahedral cavities occurs as is schematically drawn in Fig.
\ref{fig_cavities}. 
The octahedra consists of (111) pore walls. As soon as the complete surface of the octahedra reaches a 
critical value, it is easier to start a new cavity at a (100) tip of the old cavity, since the current 
density in the new, small cavity is larger and no surface passivation will occur until the surface again 
becomes to large. This growing mechanism leads to an oscillation of 
both the current through each pore 
and the diameter of each pore as a 
function of time which is also plotted in Fig. \ref{fig_cavities}. As in the case of oxide dissolution 
an internal time constant is related to the pore growth.

%\IGNFIG{
%}%IGNFIG

%
\clearpage

%\IGNFIG{
%}%IGNFIG

%Phase separation:

If a global current density smaller than the average
current density in a current burst is applied, 
the aging concept becomes essential to explain
the generation of macropores: 
At passivated surfaces, the nucleation probability
of new current bursts is lower than at sites 
where a current burst has taken place before.
Therefore the situation shown in
Fig.~\ref{fig_schem_charge_time_consuming}(c)
is much more unlikely than the situation shown in
Fig.~\ref{fig_schem_charge_time_consuming}(d),
so that the whole surface separates in two
phases with current-carrying pores
and passivated surface without contributions 
to the total current.

%\clearpage
%

\section{Consequences on Pore Geometries}
As described above, two synergetic effects 
occur as a consequence of the interaction of Current Bursts:
\begin{itemize}
\item A surface smoothing due to the  dissolution of an isotropic 
oxide layer.
\item A strongly anisotropic dissolution of the semiconductor 
due to the crystallography dependent aging of surfaces which is
one of the most important reasons for pore formation. 
\end{itemize}
Depending on the electrochemical composition of Current Bursts
the one or the other effect will dominate, and thus may lead to complete 
different surface morphologies when changing the electrochemical 
etching conditions; e.g. in a Si-HF-organic electrolyte system only by 
increasing the HF concentration (i.e. faster oxide dissolution and 
thus reduced influence of oxide) the electrochemical etching 
changes from electropolishing (strong oxide smoothing) 
over macropore formation (pores with diameters of several $\mu$m and 
smooth pore walls) to mesopore formation (strongly anisotropic, narrow 
pores with diameters of less than 400nm) 
\cite{christophersen00physstatsol}.

\begin{figure}[htbp]
\noindent
\epsfig{file=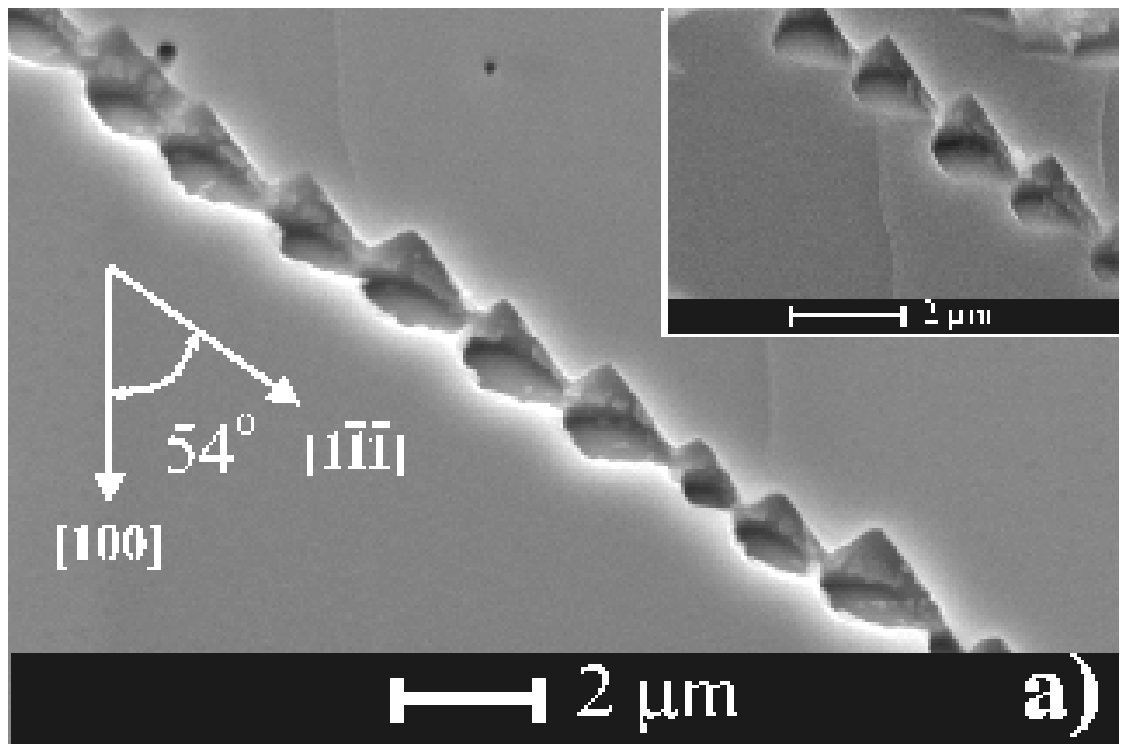,width=\columnwidth}
\\
\epsfig{file=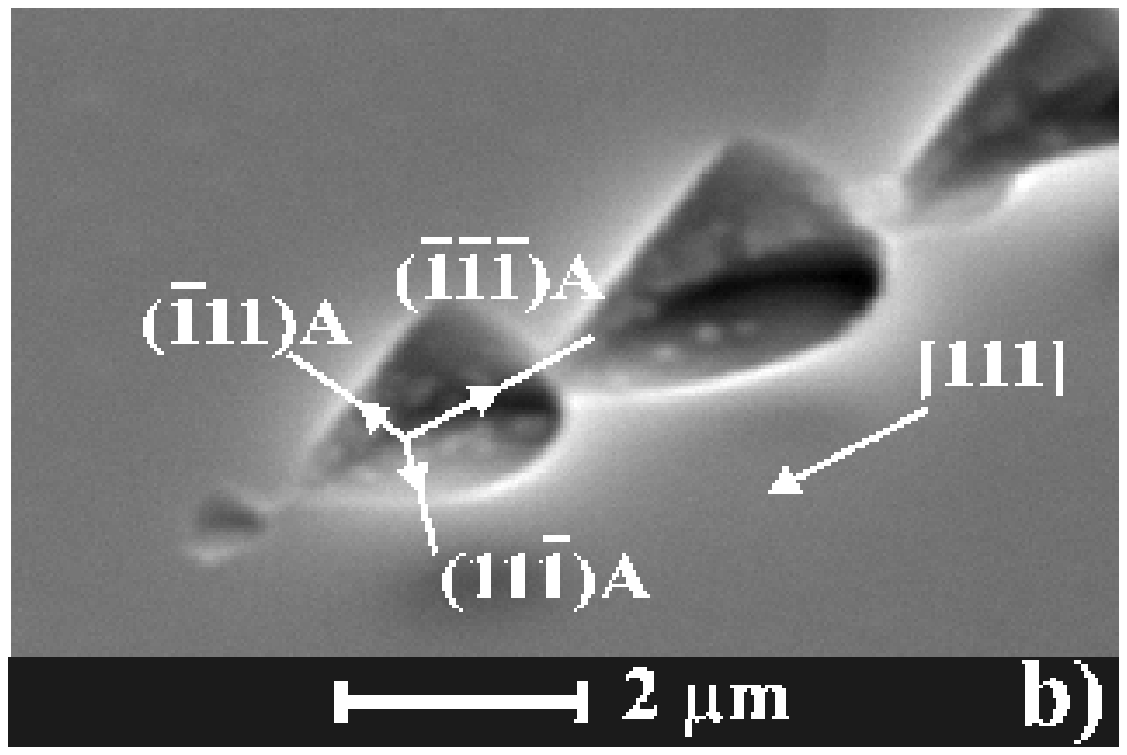,width=\columnwidth}
\caption{ \label{fig_tetrahedron}
Tetrahedron-like pores oriented along $<111>$ directions obtained in 
(100)-oriented n-GaAs ($n$ = 10$^{17}$ cm$^{-3}$)
at high current density (85 mA/cm$^2$, galvanostatic)
in HCl (5\%) electrolyte.
}
\end{figure}

Silicon has one of the most stable oxides of all semiconductors. So 
for all other semiconductors the smoothing effect is reduced, leading 
generally to rougher surfaces and smaller pore diameters. In addition the 
surface aging of III-V compounds is more complicated, since there exist two 
different (111) surfaces; e.g. in GaAs only the $\{111\}$A planes (Ga-rich 
planes) appear as stopping planes. So in most III-V compounds not 
octahedra (eight (111) surfaces as stopping planes) as in silicon are etched 
but tetrahedra with only four (111)A surfaces as stopping planes, 
and the $\{111\}$B planes (As-rich) serve as preferential growing directions 
like the (100) directions in Si (see Fig.~\ref{fig_tetrahedron}).
 
%\IGNFIG{
%}%IGNFIG

%\clearpage

\vspace{5cm}

\section{Lateral Interaction of Pores}                  
\begin{figure}[htbp]   
\epsfig{file=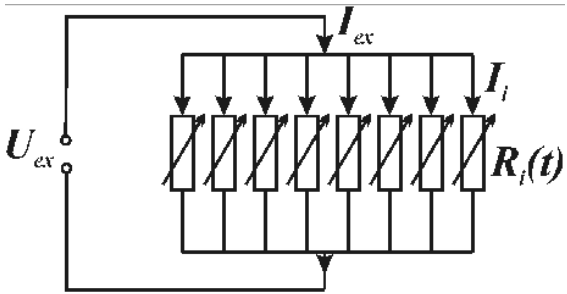,width=0.84\columnwidth} 
\caption{ \label{fig_parallelresist}
 A schematic representation of pores as oscillating resistors
in an equivalent cirquit diagram.
}
\end{figure}

\begin{figure}[htbp]
\noindent
\epsfig{file=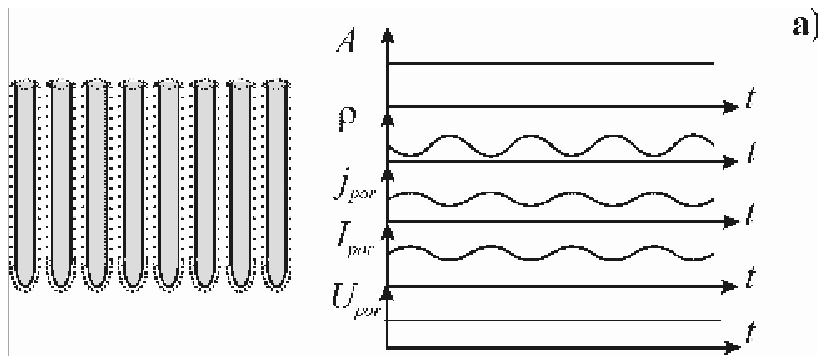,width=\columnwidth}
\\
\epsfig{file=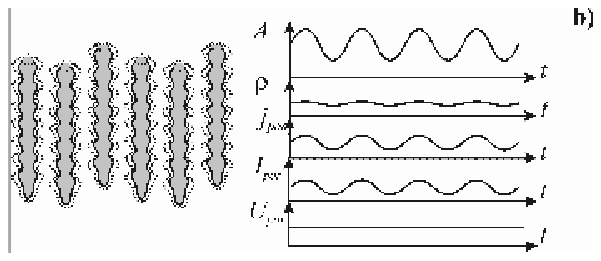,width=\columnwidth}
\\
\epsfig{file=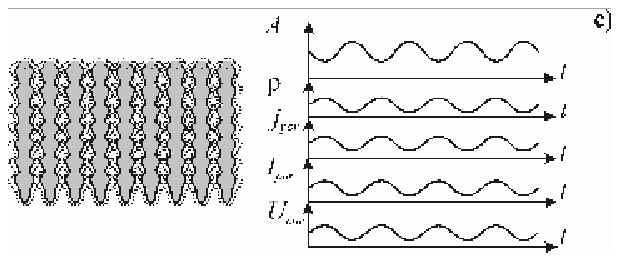,width=\columnwidth}
%\begin{tabular}[b]{c} {\scriptsize\bf d)} \\ \\\\\\\\\\\\\\\end{tabular}
\caption{ \label{fig_oszillsymb}
A schematic representation of pore interaction explaining the morphologies 
observed in Si and III-V compounds; 
a) $\rho$ oscillates and A is constant; 
b) $\rho$ is constant and A oscillates; 
c) $\rho$ and A oscillate simultaneously. 
% d) A schematic representation of pores as oscillating resistors as an equal cirquit.
Note that for galvanostatic conditions a constant total current 
is only possible for the case of random  oscillator phases.
}
\end{figure}

In the case of stable pore growth active current bursts only exist 
at each pore tip, so there is no way that current burst can interact 
directly, e.g. by overlapping. Since several types of pores grow by 
forming chains of interconnected cavities, i.e. a self induced 

diameter modulation is an intrinsic feature of such pores, they can 
interact indirectly via the space charge region between pores.
The serial resistance $R$ of each pore can be described by
\begin{equation}
R(t)=\frac{\rho(t) l(t)}{A(t)}
\end{equation}
where $\rho(t)$ is the effective specific resistance at a pore tip, 
$l(t)$ the length of the pore and $A(t)$ the chemically active area 
at the pore tip. Due to aging $\rho(t)$ will periodically increase 
and decrease while the cavities are formed. The electric circuit 
consisting of a parallel connection of large numbers of individually
oscillating pores is illustarted in Fig. \ref{fig_parallelresist}.
The consequences for the pore morphologies depend strongly on the 
etching conditions as 
schematically illustrated in Fig. \ref{fig_oszillsymb}. At constant 
voltage condition (potentiostatic control) the current through each 
pore will be modulated; since the pores grow independently of each other 
there exists no phase coupling and the overall current is constant.

\begin{figure}[htbp]
\noindent
\mbox{~}\hfill\epsfig{file=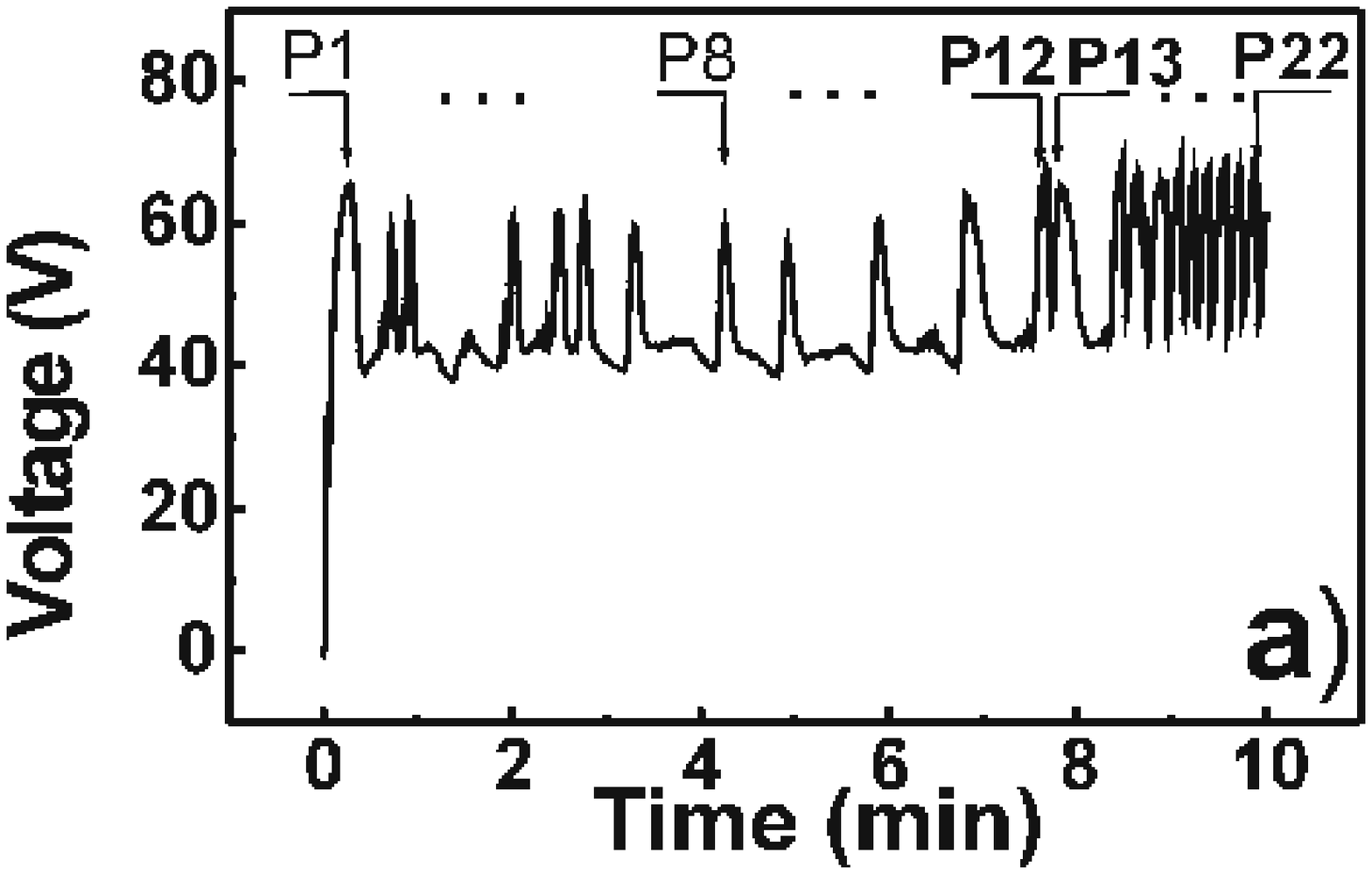,width=0.95\columnwidth}
\\
\mbox{~}\hfill\epsfig{file=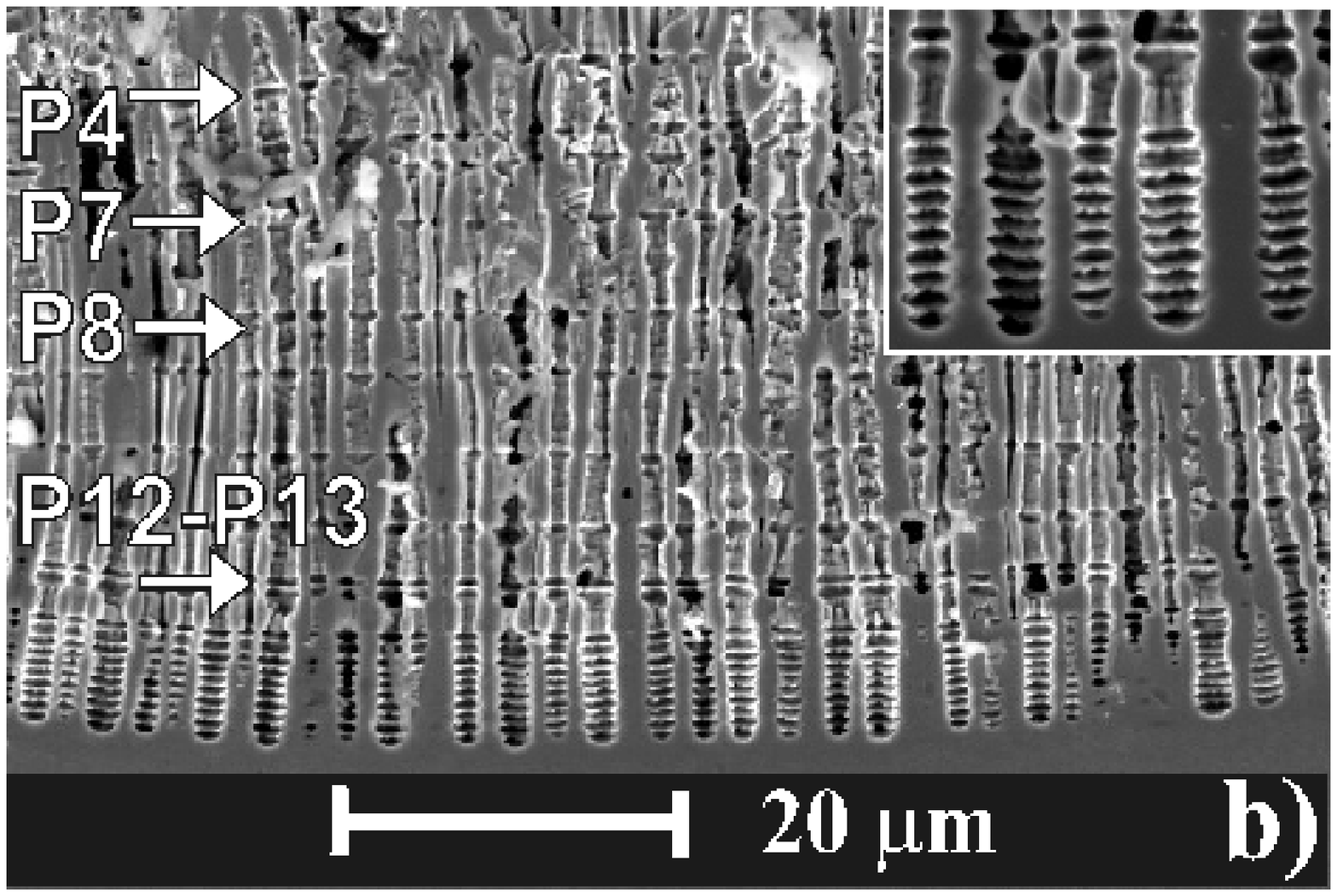,width=0.8\columnwidth}
\\
%%%%%%%%%%%%kein c%%%%\epsfig{file=image59.eps,width=0.95\columnwidth}
\\
%%% {\huge ARRAY} \vspace*{0.4\columnwidth} {~}
\mbox{~}\hfill\epsfig{file=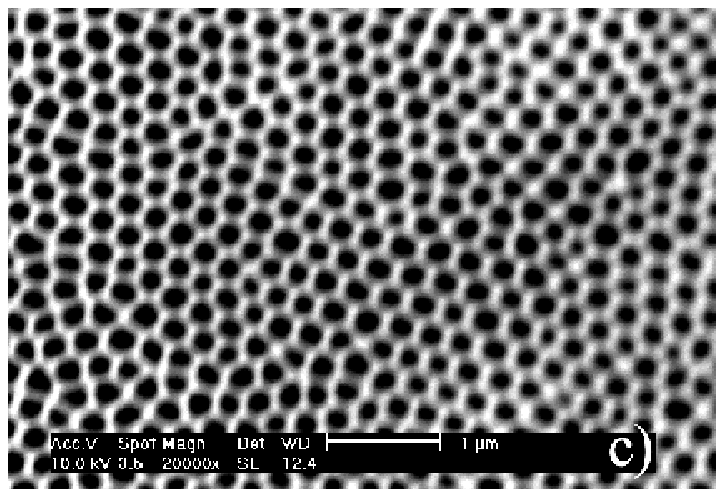,width=0.8\columnwidth}
 \caption{ \label{fig_inp_oscill}
Data taken from an (100) oriented n-InP sample with 
$n_1$ = 1.5 $\times$ 10$^{16}$ cm$^{-3}$ 
anodized at a constant current density j = 100 mA/cm$^2$. 
The electrolyte is again HCl ($5\%$).
a) Voltage oscillations; b) cross-sectional SEM of the sample. 
The inset is the magnification of the nodes at 
the bottom of the porous layer; 
%%% c) The rate of pore growth during  the anodization of InP with n = 1.5 $\times$ 10$^{16}$ cm$^{-3}$.
c) Induced by the next neighbor interaction, 
pores arrange into a close packed hexagonal 2D array.
Correlation length is about 7 lattice constants.
}
\end{figure}

For galvanostatic control we have to distinguish between two cases:
\begin{itemize}
\item At low pore densities (low external current density) there is no 
restiction for the diameter of the pores; i.e. the area $A(t)$ 
may increase freely. 
\item At high pore densities (high external current density) the 
increase of the diameter is resticted by the neighboring pores; thus 
the area $A(t)$ is limited.
\end{itemize}
In the first case we will find diameter modulation (cf. Fig. 
\ref{fig_oszillsymb} b)), but still no phase coupling between pores.
In the second case the area of the pores can cot increase freely to 
compensate for the reduced effective specific resistance $\rho(t)$.
Thus the external voltage must increase to guarantee a constant external 
current. The voltage increase should synchronize the phases of growth 
for all pores. This is exactly what is found when, e.g., etching InP at 
very high current densities. In Fig. \ref{fig_inp_oscill} a) self 
induced voltage oscillations are shown which coincide with a simultaneous 
diameter increase of all pores (cf. the cross section of the pores in 
Fig. \ref{fig_inp_oscill} b)). Obviously not all cycles of the pore growth have 
been synchronized globally allthough nearly a closed packed ordering 
of the pores was reached (cf. plane view of pore array in Fig. 
\ref{fig_inp_oscill} c)).

%\IGNFIG{
%}%IGNFIG

% \IGNFIG{
%}%IGNFIG

%\clearpage

\vspace{10cm}

\section{Open-Loop-Control of Dynamical Systems}
Contrary to the methods of Ott, Grebogi and Yorke \cite{ogy90}
and of Pyragas \cite{pyragas}, which allow for the stabilization
of (if uncontrolled) unstable periodic orbits by 
parametric feedback, it may be possible to achieve 
stabilization even without feedback, i.~e. 
by open-loop control.
Application of non-feedback control to dynamical 
systems can be roughly classified into at least three
classes:
Addition of noise \cite{malescioPLA,lythe}
or of chaotic signals \cite{rajasekar}
may lead to suppression of chaos.
Second, a constant (but usually large) shift in a parameter 
may be utilized to shift a fixed point into a 
parameter regime where it is stable \cite{parth}.
Third, periodic parametric perturbations can 
be used. Counterintuitively, this is possible
even by nonresonant perturbations
\cite{kivsharbenner}. 
The more straightforward approach is by
resonant parametric perturbations 
\cite{lima90,filatrella,mettinpre,mettin,rhodes,sukow,colet}
which in the simplest case is a weak
sinusoidal signal, but can be any periodic signal.
Which orbits can be stabilized by which signals
in general can be investigated only 
if an explicit model of the system is given.
\section{Open-Loop-Control of Pore Formation}
\begin{figure}[htbp]  \noindent
\centerline{
\epsfig{file=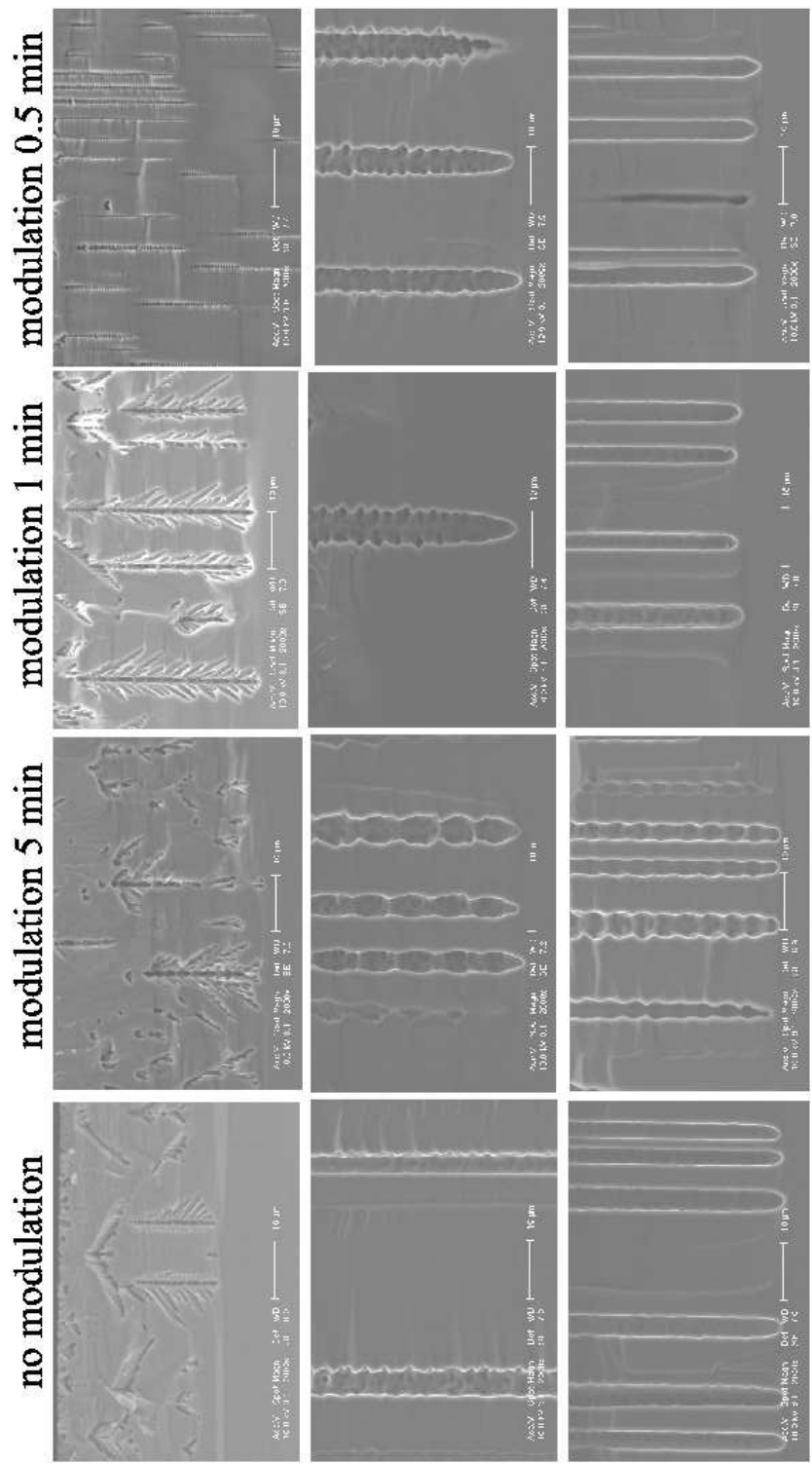,%
%height=0.8\columnwidth,%
width=0.78\textheight,%
angle=90}
%
%\mbox{~}\\ \vspace*{4cm}\mbox{~}\\  {LARGE Openloop-Bild!!!! grosse Datei!!!} \vspace*{4cm} {~}
}
\caption{ \label{fig_openloop}
Branching of pores can be suppressed by 
an open-loop control method triggering to the
system-inherent time scales (here: (100)-oriented n-Si, 1-10$\Omega$cm).
The upper row (DMF, 4wt-\% HF)
corresponds to the case with less oxide generation.
From middle row (H$_2$O, 7wt-\% HF)
to lower row (H$_2$O, 4wt-\% HF), there is 
less oxide dissolution. 
In the first column the 
backside illumination current is constant; the triggering 
time decreases from 5 min in the second colume to 0.5 min 
in the last column. 
Here potentiostatic conditions (4V) are used;
a 20$\%$ sinusoidal modulation 
(with the respective period)
of the current density around its average value of
4 mA/cm$^2$ is realized by control of the 
amplitude of backside illumination.
}
\end{figure}

For technical applications, it would be desirable to
generate pores with a diameter of less than 500nm.
However, at all etching conditions investigated up to 
now, in this diameter range always side-branching pores 
are obtained (cf. first row in Fig. \ref{fig_openloop}). 

Fortunately, the experimental setup allows for an open-loop 
stabilization e.g. by modulation of the intensity of 
illumination or by modulation of the etching current. 
If the modulation frequency meets the intrinsic time scale
of the pore formation (0.5 min modulation in row 1), 
the side pore formation is hindered 
by the reduction of the etching current at the right phase.
Which is the relevant time scale (oxide dissolution time, 
passivation time, ...) depends on the etching conditions 
and the type of pores. 

Corresponding to the modulated current, 
the pores are regular, but with a modulated
diameter (Fig.~\ref{fig_openloop}). 
The diameter modulation profile can be
influenced to some extent by the waveform, 
nota bene these waveforms are not identical. 

\typeout{Fig. fig_openloop goes here}

It should be noted, although the controlled state 
possibly could be considered as an instable
periodic orbit
of the whole etching setup,
the described control method yet does not use a feedback
control scheme as the OGY scheme \cite{ogy90}.
This would be a promising extension,
however, it is expected that for the case 
of dendritic-like sidebranching, 
the dynamics possibly cannot be reduced to a low-dimensional
attractor.
 
\clearpage
%\IGNFIG{
%
%
\onecolumn
%}%IGNFIG

\clearpage
\twocolumn
\section{Conclusions}
The Current Burst model, a local stochastic
and highly nonlinear model, and the Aging concept 
describing the time- and orientation-dependence
of the passivation behavior,
provide on an intermediate level of abstraction 
a general approach to explain pore geometries,
oscillations and synchronization
with a minimum of material-dependent,
but experimentally accessible, parameters.
The lateral interaction of Current Bursts gives rise to 
synchronization phenomena,
and a percolation transition to global ordering.
In a technologically relevant diameter range,
where one otherwise experiences excessive sidebranching 
of pores giving dendritic-like structures,
an open-loop control can be successfully applied to suppress
sidebranching of pores. 
To clarify quantitatively the dynamical processes 
in this system, and to gain detailed insight
about the general controllability conditions 
for mesopores and perspectives for 
more sophisticated control strategies 
further experimental and theoretical
efforts have to be made.

% 
%\clearpage

%
% ---- Bibliography ----
%

%
\end{document}